\documentclass[reprint,amsmath,amssymb,aps,prb,showpacs]{revtex4-1}
\usepackage{graphicx}
\usepackage{graphics}
\begin{document}
\title{Heat transport in an anharmonic crystal}
\author{Shiladitya Acharya}
\author{Krishnendu Mukherjee}
\email{kmukherjee@physics.iiests.ac.in}
\affiliation{Department of Physics, Indian Institute of Engineering Science
and Technology, Shibpur, Howrah- 711103, West Bengal, India}

\begin{abstract}
We take an ordered, anharmonic crystal in the form of slab geometry
in three dimensions. Apart from attaching baths of Langevin type 
to the extreme surfaces, we also attach baths of same 
type to the intermediate surfaces of the slab to simulate the 
environment surrounding the system. 
We assume noise functions to be Gaussian and 
their widths to be site dependent. We find that the 
radiated heat from the slab does not receive any correction at the 
leading order of anharmonic coupling and the Newton's law of cooling
holds for an appropriate choice of the widths. We observe that in the 
steady state limit entire slab becomes an assembly of $N$ different thermally
equilibriated layers, where $N$ is the number of sites in the direction
of conduction current flow. We find an exponentially
falling nature of the temperature profile as its leading behaviour
and its non-leading behaviour is governed by the two site 
dependent functions. 
Our evaluation suggests that in the thermodynamic limit thermal
conductivity remains independent of the environment temperature 
and is dependent only on the difference of temperature of the 
extreme surfaces linearly at the leading order of anharmonic coupling.  
We find that owing to finiteness of conductivity in the thermodynamic limit,
Fourier's law holds to leading order in anharmonic coupling.
 
\end{abstract}
\maketitle

\section{Introduction}

The process of heat transport
in a solid mainly involves conduction, radiation and absorption 
of heat. When the absorption is absent in the steady state limit, the 
temperature profile of a solid bar falls exponentially from high to 
low temperature end of the bar if
the radiated heat obeys Newton's laws of cooling
and the conducted current density obeys Fourier's law:
\begin{equation}
\mathbf{J}(\mathbf{x})=-\kappa{\mathbf \nabla}T(\mathbf{x}), 
\end{equation}
where ${\mathbf \nabla}T(\mathbf{x})$ is the local temperature gradient
and $\kappa$ is the thermal conductivity.
In the steady state the exponentially
falling nature of the temperature profile is
experimentally verified by Ingen and Hausz\cite{Saha1969}.
Through ages it is a challenge to theorist\cite{Bonetto2000, Lepri2003} 
to derive the laws involved with this process from the application of 
basic principles of statistical mechanics to a crystal of solid.

Our study is based on Langevin equation approach\cite{Chaikin2009,
Reif1985} which was first used to study heat conduction in ordered,
harmonic lattice in one-dimension\cite{Rieder1967}.
Later, this approach was used to study heat conduction in regular
harmonic and anharmonic lattices in one, two and three
dimensions\cite{Nakazawa1970}.
This approach was also used to study heat conduction in disordered,
harmonic system in one\cite{Rubin1971,Connor1974,Dhar2001} and
two\cite{Lee2005} dimensions. This approach was adopted in disordered,
harmonic system in two and three dimensions\cite{Chaudhuri2010} and 
the validity of Fourier's law was numerically verified in pinned three
dimensional system. The Fourier's law was also verified numerically 
using this approach in three dimensional anharmonic crystal\cite{Saito2010}.
Unlike the model of self consistent reservoirs\cite{Bolsterli1970,
Rich1975,Bonetto2004}, if one allows the heat to flow between 
intermediate surfaces and the attached baths, 
one derives Fourier's law in three dimensional 
ordered harmonic system using this approach and obtains the 
exponentially falling temperature profile, 
provided that the radiated heat obeys the Newton's law of 
cooling\cite{Shila2013}. 

In this paper we take the ordered, anharmonic crystal in three dimension
in the form of slab geometry.
Apart from the baths attached to the extremities, there are baths 
attached to the intermediate surfaces of the slab and the heat is 
allowed to flow between intermediate surfaces and the attached 
baths\cite{Shila2013}. We show using Langevin equation approach 
that Fourier's law remains valid in the thermodynamic limit and 
a temperature dependent conductivity is obtained to leading order 
in anharmonic coupling. We observe that the radiated heat does not
receive any correction form the anharmonic part. We obtain the 
exact temperature profile which is correct to leading order in 
anharmonic coupling and observe that the exponentially falling
nature of the profile is modified by the two site 
dependent functions at the leading order in anharmonic coupling.

We organize the paper as follows. 
We discuss the model and obtain the steady state solution
of the Langevin equation perturbatively to leading order in 
anharmonic coupling in section II. We compute the correlators 
in section III. We compute the radiated heat and obtain the 
temperature profile in section IV. We compute the conduction 
current density and obtain the Fourier's law in section V. 
We summarize our results in section VII.     
We give a brief outline of the evaluation of frequency
integrals in appendix A. We discuss the wave vector sums 
in the continuum limit in appendix B. We give the definition 
of the useful integrals and the required discrete sums in 
appendices C and D respectively.

\section{Model and its solution}

We consider an anharmonic crystal in three dimensions and each
lattice site ${\mathbf n}=(n_1,\,n_2,\,n_3)$ is described by the 
three integers in the regions: $1\le n_1\le\,N$, $1\le n_2\le\, W_2$
and $1\le n_3\le\, W_3$. We attach Langevin type heat baths with
the surfaces at $n_1=1$ and $n_1=N$ maintaining fixed temperatures
$T_1$ and $T_N$ ($T_1\,>\,T_N$) respectively. Since the crystal is
exposed to the surroundings, the mass points of the lattice  
interact with the environment. To simulate this interaction we also 
attach heat baths of the same type to the intermediate surfaces from
$n_1=2$ to $n_1=N-1$. If $x_{{\mathbf n}}$ be the displacement field  
of a lattice site, the Lagrangian
\begin{equation}
L=m\sum_{\mathbf n}\frac{\dot{x}_{\mathbf n}^2}{2}
-\sum_{{\mathbf n},\,\hat{e}}\Big\{
\frac{C_2}{4}(x_{{\mathbf n}+\hat{e}}-x_{{\mathbf n}})^2
+\frac{C_4}{8}(x_{{\mathbf n}+\hat{e}}-x_{{\mathbf n}})^4\Big\},
\end{equation} 
where $m$ is the mass attached to each lattice point. $C_2$ and
$C_4$ are the force constants for harmonic and quartic interactions 
respectively. $\hat{e}$ denotes the unit vector in three dimension.
In presence of the heat baths the equation of motion of a mass point
at the lattice site ${\mathbf n}$ reads
\begin{eqnarray}
m\ddot{x}_{\mathbf n} &=&
-\sum_{\hat{e}}\{C_2(x_{\mathbf n}-x_{{\mathbf n}+\hat{e}})
+C_4(x_{\mathbf n}-x_{{\mathbf n}+\hat{e}})^3\}\nonumber\\
& &-m\gamma\dot{x}_{\mathbf n}
+\eta_{\mathbf n},
\label{eom1}
\end{eqnarray} 
where $\gamma$ denotes the damping constant (or coefficient of friction)
of the medium and $\eta_{\mathbf n}$ denotes the noise (or random force)
offered by the baths. We choose the probability distribution of noise
Gaussian and is given by
\begin{equation}
P[\eta]=\exp\Big[-
\sum_{\mathbf n}\frac{1}{2m\gamma z_{n_1}}\int_{-\infty}^\infty dt
\,\eta_{\mathbf n}^2(t)\Big],
\label{noise}
\end{equation} 
where $\gamma\,> 0$.
We have chosen the width $\sqrt{ m\gamma z_{n_1}}$ of the 
distribution dependent on $n_1$. We shall determine $z_{n_1}$
using fluctuation dissipation theorem in the steady state 
limit. According to the choice of eq.(\ref{noise}), correlations
of odd number of noise functions vanish. Correlations 
of even number of noise functions are non zero and can be 
expressed in terms of two point correlation function
\begin{eqnarray}
\langle \eta_{\mathbf n}(t) 
 \,\eta_{{\mathbf n}^\prime}(t^\prime)\rangle 
&=&\frac{\int{\cal D}\eta\, P[\eta]\,
\eta_{\mathbf n}(t)\,\eta_{{\mathbf n}^\prime}(t^\prime)}
{\int{\cal D}\eta\, P[\eta]}\nonumber\\
&=& m\gamma z_{n_1}\delta(t-t^\prime)\delta_{{\mathbf n},{\mathbf n}^\prime}, 
\label{etacorrelation}
\end{eqnarray}
where the dimension of $z_{n_1}$ is same as the dimension of
energy. We use the periodic boundary conditions in $n_2$ and
$n_3$ directions for the displacement field and noise:
\begin{eqnarray}
x_{{\mathbf n}+(0,W_2,0)}(t) &=& x_{\mathbf n}
= x_{{\mathbf n}+(0,0,W_3)}(t),\nonumber\\
\eta_{{\mathbf n}+(0,W_2,0)}(t) &=& \eta_{\mathbf n}
= \eta_{{\mathbf n}+(0,0,W_3)}(t).
\end{eqnarray}
These conditions lead to the Fourier expansions:
\begin{eqnarray}
x_{\mathbf n}(t) &=&
\frac{1}{\sqrt{W_2W_3}}\sum_{\mathbf p}
y_{n_1}({\mathbf p},t) e^{ia{\mathbf p}.{\mathbf n}_\perp},
\label{fexnt}\\
\eta_{\mathbf n}(t) &=&
\frac{1}{\sqrt{W_2W_3}}\sum_{\mathbf p}
f_{n_1}({\mathbf p},t) e^{ia{\mathbf p}.{\mathbf n}_\perp},
\label{feetant}
\end{eqnarray}
where $a$ is the lattice constant, ${\mathbf p}=(p_2,\, p_3)$
and ${\mathbf n}_\perp=(n_2,\, n_3)$. 
We use eq.(\ref{feetant}) in eq.(\ref{etacorrelation})
to obtain the two point correlation of Fourier transformed
noise functions as
\begin{equation}
\langle f_j({\mathbf p},t)f_k({\mathbf p}^\prime,t^\prime)\rangle
=m\gamma z_j\delta_{jk}\delta(t-t^\prime)
\delta_{{\mathbf p}+{\mathbf p}^\prime,0}
\label{fft}
\end{equation}
We then use eq.(\ref{fexnt})
and (\ref{feetant}) in eq.(\ref{eom1}) to obtain
\begin{eqnarray}
\ddot{y}_j({\mathbf q},t) &=&
-\frac{\omega_h^2}{4}
\sum_{k=1}^NV_{jk}({\mathbf q})y_k({\mathbf q},t)
-\gamma\dot{y}_j({\mathbf q},t)\nonumber\\
& &+\frac{1}{m}f_j({\mathbf q},t)
-F_j({\mathbf q},t),
\label{eom2}
\end{eqnarray}
where $\omega_h^2=\frac{4C_2}{m}$, the $N\times N$ matrix
\begin{equation}
V({\mathbf q})=\left(\begin{array}{ccccc}
2\omega_0^2 & -1 & 0 & 0 & \ldots  \\ 
-1 & 2\omega_0^2 & -1 & 0 & \ldots  \\
0 & -1 & 2\omega_0^2 & -1 & \ddots \\
\vdots & \ddots & \ddots & \ddots & \ddots  \\ 
0 & \ldots & 0 & -1 & 2\omega_0^2
\end{array}\right)
\label{Vmatrix}
\end{equation}
and
\begin{equation} 
\omega_0^2({\bf{q}}) = 1+2\sin^2(\frac{q_2 a}{2})
+2\sin^2(\frac{q_3 a}{2}).
\label{w0}
\end{equation}
The dimensionless quartic coupling constant
\begin{eqnarray}
\lambda &=& \frac{4a^2C_4}{m\omega_h^2},
\label{lambda}\\
V_4({\mathbf p},{\mathbf p}^\prime,{\mathbf p}^{\prime\prime})
&=& 16\Big[\sin(p_2a/2)\sin(p_2^\prime a/2)
\sin(p_2^{\prime\prime}a/2)\nonumber\\
& &\times\sin((p_2+p_2^\prime+p_2^{\prime\prime})a/2)\nonumber\\
& &+\sin(p_3a/2)\sin(p_3^\prime a/2)
\sin(p_3^{\prime\prime}a/2)\nonumber\\
& &\times\sin((p_3+p_3^\prime+p_3^{\prime\prime})a/2)\Big]
\label{V4}
\end{eqnarray}
and
\begin{eqnarray}
F_j({\mathbf q},t) &=& \frac{\lambda\omega_h^2}{4a^2 W_2 W_3}
\sum_{{\mathbf p},{\mathbf p}^\prime}\Big\{
(y_j({\mathbf p},t)-y_{j+1}({\mathbf p},t))\nonumber\\
& &\times(y_j({\mathbf p}^\prime,t)-y_{j+1}({\mathbf p}^\prime,t))\nonumber\\
& &\times(y_j({\mathbf q}-{\mathbf p}-{\mathbf p}^\prime,t)
-y_{j+1}({\mathbf q}-{\mathbf p}-{\mathbf p}^\prime,t))\nonumber\\
& &+
(y_j({\mathbf p},t)-y_{j-1}({\mathbf p},t))
(y_j({\mathbf p}^\prime,t)-y_{j-1}({\mathbf p}^\prime,t))\nonumber\\
& &\times(y_j({\mathbf q}-{\mathbf p}-{\mathbf p}^\prime,t)
-y_{j-1}({\mathbf q}-{\mathbf p}-{\mathbf p}^\prime,t))\nonumber\\
& &-V_4({\mathbf p},{\mathbf p}^\prime,{\mathbf q}-{\mathbf p}
-{\mathbf p}^\prime) y_j({\mathbf p},t) y_j({\mathbf p}^\prime,t)\nonumber\\
& &\times y_j({\mathbf q}-{\mathbf p}-{\mathbf p}^\prime,t)\}.
\label{Fj}
\end{eqnarray}
We have assumed that $y_0({\mathbf p}, t)=0=y_{N+1}({\mathbf p},t)$.
We choose a coordinate system where  
the matrix $V({\mathbf q})$ is diagonal and this has been 
accomplished using an orthogonal matrix $A$ such that 
$A^TV({\mathbf q})A=\alpha^2({\mathbf q})$, where 
$(\alpha^2({\mathbf q}))_{jk}=\alpha^2_j({\mathbf q})\delta_{jk}$,
\begin{equation}
\alpha_j^2({\mathbf q})=2\omega_0^2({\mathbf q})
+2\cos j\nu, ~~\nu=\pi/(N+1)
\end{equation}
and 
\begin{equation}
A_{j,k}=\sqrt{\frac{1}{N+1}}(-1)^{j+1}\sin(jk\nu).
\label{Ajk}
\end{equation}  
In terms of new set of coordinates $\xi_j({\mathbf p},t)$ 
($j=1, \cdots, N$) which is defined by the equation
\begin{equation}
y_j({\mathbf q},t)=\sum_{k=1}^N A_{j,k}\xi_k({\mathbf q},t),
\label{yxi}
\end{equation}
eq.(\ref{eom2}) reads
\begin{eqnarray}
\ddot{\xi}_j({\mathbf q},t) &=&
-\omega_j^2({\mathbf q})\xi_j({\mathbf q},t)
-\gamma\dot{\xi}_j({\mathbf q},t)
+\frac{1}{m}\tilde{f}_j({\mathbf q},t)\nonumber\\
& &-\tilde{F}_j({\mathbf q},t),
\label{eom3}
\end{eqnarray}
where
\begin{eqnarray}
\omega_j^2({\mathbf q}) &=& \frac{\omega_h^2}{4}
\alpha_j^2({\mathbf q})
=\frac{\omega^2_h}{2}(\omega^2_0({\mathbf q})
+\cos j\nu),\label{wjq}\\
\tilde{f}_j({\mathbf q},t) &=& \sum_{k=1}^N 
A_{k,j}f_k({\mathbf q},t),\label{ftilde}\\
\tilde{F}_j({\mathbf q},t) &=& \sum_{k=1}^N 
A_{k,j}F_k({\mathbf q},t)\nonumber\\
&=&\frac{\lambda\omega_h^2}{4a^2 W_2 W_3}
\sum_{{\mathbf p},{\mathbf p}^\prime}
\sum_{k_1,k_2,k_3=1}^N X_{jk_1k_2k_3}
({\mathbf q}, {\mathbf p},{\mathbf p}^\prime)\nonumber\\
& &\times\xi_{k_1}({\mathbf p}, t)
\xi_{k_2}({\mathbf p}^\prime, t)
\xi_{k_3}({\mathbf q}-{\mathbf p}-{\mathbf p}^\prime, t).
\label{Ftilde}
\end{eqnarray}
The expression
\begin{eqnarray}
& & X_{jk_1k_2k_3}
({\mathbf q}, {\mathbf p},{\mathbf p}^\prime)
=\sum_{k=1}^N A_{k,j}\Big\{(A_{k,k_1}-A_{k+1,k_1})\nonumber\\
& &\times(A_{k,k_2}-A_{k+1,k_2})
(A_{k,k_3}-A_{k+1,k_3})\nonumber\\
& &+(A_{k,k_1}-A_{k-1,k_1})
(A_{k,k_2}-A_{k-1,k_2})\nonumber\\
& &\times(A_{k,k_3}-A_{k-1,k_3})\nonumber\\
& &-V_4({\mathbf p},{\mathbf p}^\prime,{\mathbf q}-{\mathbf p}
-{\mathbf p}^\prime)A_{k,k_1}A_{k,k_2}A_{k,k_3}\Big\}.
\end{eqnarray}
We then use eq.(\ref{Ajk}) and evaluate the sum over $k$. Final
expression takes the following form
\begin{eqnarray}
& & X_{jk_1k_2k_3}
({\mathbf q}, {\mathbf p},{\mathbf p}^\prime)\nonumber\\
&=&\frac{8N}{(N+1)^2}\sum_{s,s_2,s_3=\pm} ss_2s_3
\times\delta_{k_1+s_2k_2+s_3k_3+sj,0}\nonumber\\
& &\times\Big\{\cos\frac{k_1\nu}{2}\cos\frac{k_2\nu}{2}
\cos\frac{k_3\nu}{2}\cos\frac{j\nu}{2}\nonumber\\
& &-\frac{1}{16}V_4({\mathbf p},{\mathbf p}^\prime,
{\mathbf q}-{\mathbf p}-{\mathbf p}^\prime)\Big\}.
\label{X}
\end{eqnarray}

We solve eq.(\ref{eom3}) taking $\lambda$ ($<<\,1$) as a perturbation 
parameter. We set $\lambda =0$ in the eq.(\ref{eom3}) to obtain the 
zero-th order solution $\xi^{(0)}_j({\mathbf q},t)$ which satisfies
the equation
\begin{equation}
\ddot{\xi}^{(0)}_j({\mathbf q},t) =
-\omega_j^2({\mathbf q})\xi^{(0)}_j({\mathbf q},t)
-\gamma\dot{\xi}^{(0)}_j({\mathbf q},t)
+\frac{1}{m}\tilde{f}_j({\mathbf q},t).
\label{0order}
\end{equation}
It is an equation of a damped oscillator under the influence
of a random force. In the steady state ($t>>1/\gamma$) 
particular solution of the equation dominates over the 
complementary solution. Upon substitution of temporal
Fourier transform of
\begin{eqnarray}
& &~~\xi^{(0)}_j({\mathbf q},t)
=\int_{-\infty}^\infty\frac{d\omega}{2\pi} e^{i\omega t}  
\xi^{(0)}_j({\mathbf q},\omega)\\
&{\rm and}&
~~f_j({\mathbf q},t)
=\int_{-\infty}^\infty\frac{d\omega}{2\pi} e^{i\omega t}  
f_j({\mathbf q},\omega)
\end{eqnarray}
into eq.(\ref{0order}), we obtain
\begin{equation}
\xi^{(0)}_j({\mathbf q},\omega)
=\frac{1}{m}D_j({\mathbf q},\omega)\tilde{f}_j({\mathbf q},\omega),
\label{xi0w}
\end{equation}
where
\begin{equation}
D_j({\mathbf q},\omega)=-\frac{1}{\omega^2-\omega_j^2({\mathbf q})
-i\gamma\omega}
\end{equation}
To obtain the equation of motion next to leading order we set 
\begin{equation}
\xi_j({\mathbf q},t)
=\xi^{(0)}_j({\mathbf q},t)+\xi^{(1)}_j({\mathbf q},t)
\label{xit}
\end{equation} 
in eq.(\ref{eom3}) where the first order solution 
$\xi^{(1)}_j({\mathbf q},t)\sim O(\lambda)$.
Since $F_j({\mathbf q},t)\sim O(\lambda)$,
the equation of motion reads
\begin{equation}
\ddot{\xi}^{(1)}_j({\mathbf q},t)
=-\omega^2_j({\mathbf q})\xi^{(1)}_j({\mathbf q},t)
-\gamma\dot{\xi}^{(1)}_j({\mathbf q},t)-\tilde{F}^{(0)}_j({\mathbf q},t),
\label{1order}
\end{equation} 
where $\tilde{F}^{(0)}_j({\mathbf q},t)$ is obtained 
from the expression of $\tilde{F}({\mathbf q},t)$ setting 
$\xi_j({\mathbf q},t)=\xi^{(0)}_j({\mathbf q},t)$ in eq.(\ref{Ftilde}).
Using
\begin{equation}  
\xi^{(1)}_j({\mathbf q},t)
=\int_{-\infty}^\infty\frac{d\omega}{2\pi}
e^{i\omega t}
\xi^{(1)}_j({\mathbf q},\omega),
\end{equation}
in eq.(\ref{1order}), we obtain in the steady state 
\begin{equation}
\xi^{(1)}_j({\mathbf q},\omega)
=-D_j({\mathbf q},\omega)
\tilde{F}^{(0)}_j({\mathbf q},\omega),
\label{xi1w}
\end{equation} 
where
\begin{eqnarray}
\tilde{F}^{(0)}_j({\mathbf q},\omega)
&=&\frac{\lambda\omega_h^2}{4a^2W_2W_3}
\sum_{{\mathbf q}^\prime,{\mathbf q}^{\prime\prime}}
\sum_{k_1,k_2,k_3=1}^N
X_{jk_1k_2k_3}({\mathbf q},{\mathbf q}^\prime,{\mathbf q}^{\prime\prime})
\nonumber\\
& &\times\int_{-\infty}^\infty\frac{d\omega^\prime}{2\pi}
\frac{d\omega^{\prime\prime}}{2\pi}
\xi^{(0)}_{k_1}({\mathbf q}^\prime,\omega^\prime)
\xi^{(0)}_{k_2}({\mathbf q}^{\prime\prime},\omega^{\prime\prime})\nonumber\\
&&\times\xi^{(0)}_{k_3}
({\mathbf q}-{\mathbf q}^\prime-{\mathbf q}^{\prime\prime},
\omega-\omega^\prime-\omega^{\prime\prime}).
\label{Ftildew}
\end{eqnarray}

\section{Correlators}

We use the temporal Fourier transform of noise functions in 
eq.(\ref{fft}) and obtain the noise correlation in frequency
space as
\begin{equation}
\langle f_j({\mathbf q},\omega)
f_k({\mathbf q}^\prime,\omega^\prime)\rangle
=2\pi\gamma mz_j\delta_{jk}
\delta_{{\mathbf q}+{\mathbf q}^\prime,0}
\delta(\omega+\omega^\prime).
\label{ffw}
\end{equation}
With the use of this correlator and the solution in eq.(\ref{xi0w})
we obtain the correlation function between zero-th order displacement 
fields in frequency space
\begin{equation}
\langle\xi^{(0)}_j({\mathbf q},\omega)
\xi^{(0)}_k({\mathbf q}^\prime,\omega^\prime)\rangle
= G^{(0)}_{jk}({\mathbf q}, \omega)
\delta_{{\mathbf q}+{\mathbf q}^\prime,0}
\delta(\omega+\omega^\prime),
\label{xi00w}
\end{equation}
where
\begin{eqnarray}
G^{(0)}_{jk}({\mathbf q}, \omega)
&=&\frac{2\pi\gamma}{m}\sum_{l=1}^N
z_l A_{l,j} A_{l,k} D_j({\mathbf q},\omega)\nonumber\\
& &\times D_k(-{\mathbf q},-\omega).
\label{G0}
\end{eqnarray}
We use eq.(\ref{xi1w}) to obtain the  correlation function 
between zeroth and first order displacement fields as
\begin{eqnarray}
& &\langle\xi^{(0)}_j({\mathbf q},\omega)
\xi^{(1)}_k({\mathbf q}^\prime,\omega^\prime)\rangle\nonumber\\
&=&-D_k({\mathbf q}^\prime, \omega^\prime)
\langle\xi^{(0)}_j({\mathbf q},\omega)
F^{(0)}_k({\mathbf q}^\prime,\omega^\prime)\rangle\nonumber\\
&=&-\frac{\lambda\omega_h^2}{4a^2W_2W_3}
\sum_{{\mathbf q}^{\prime\prime},{\mathbf q}^{\prime\prime\prime}}
\sum_{j_1,j_2,j_3=1}^N
X_{kj_1j_2j_3}({\mathbf q}^\prime,{\mathbf q}^{\prime\prime},
{\mathbf q}^{\prime\prime\prime}) \nonumber\\
& &\times\int_{-\infty}^\infty\frac{d\omega^{\prime\prime}}{2\pi}
\frac{d\omega^{\prime\prime\prime}}{2\pi}
\langle\xi^{(0)}_j({\mathbf q},\omega)
\xi^{(0)}_{j_1}({\mathbf q}^{\prime\prime},\omega^{\prime\prime})
\xi^{(0)}_{j_2}({\mathbf q}^{\prime\prime\prime},
\omega^{\prime\prime\prime})\nonumber\\
& &\times\xi^{(0)}_{j_3}
({\mathbf q}^\prime-{\mathbf q}^{\prime\prime}
-{\mathbf q}^{\prime\prime\prime},
\omega^\prime-\omega^{\prime\prime}-\omega^{\prime\prime\prime})\rangle,
\end{eqnarray}
where we have used eq,(\ref{Ftildew}) in the last step.
The four point correlation of displacement fields can 
be decomposed in terms of two point correlation functions as
\begin{eqnarray}
& &\langle\xi^{(0)}_j({\mathbf q},\omega)
\xi^{(1)}_k({\mathbf q}^\prime,\omega^\prime)\rangle\nonumber\\
&=&-\frac{\lambda\omega_h^2}{8\pi a^2W_2W_3}
D_k{(\mathbf q^{\prime},\omega^{\prime})}
\delta_{{\mathbf q}+{\mathbf q}^\prime,0}
\delta(\omega+\omega^{\prime})\nonumber\\
& &\times\sum_{j_1,j_2,j_3=1}^N\sum_{\mathbf q^{\prime\prime}}
\int_{-\infty}^{\infty}
\frac{d\omega^{\prime\prime}}{2\pi}\nonumber\\
& &\times\Big[X_{kj_1j_2j_3}({\mathbf q}^\prime,{\mathbf q}^\prime,
{\mathbf q}^{\prime\prime})G^{(0)}_{jj_1}({\mathbf q},\omega)
G^{(0)}_{j_2j_3}({\mathbf q}^{\prime\prime},\omega^{\prime\prime})\nonumber\\
&&+ X_{kj_1j_2j_3}({\mathbf q}^{\prime},{\mathbf q}^{\prime\prime},
{\mathbf q}^{\prime}) G^{(0)}_{jj_2}({\mathbf q},\omega)
G^{(0)}_{j_1j_3}({\mathbf q}^{\prime\prime},\omega^{\prime\prime})\nonumber\\
&&+ X_{kj_1j_2j_3}({\mathbf q}^{\prime},{\mathbf q}^{\prime\prime},
-{\mathbf q}^{\prime\prime})G^{(0)}_{jj_3}({\mathbf q},\omega)
G^{(0)}_{j_1j_2}({\mathbf q}^{\prime\prime},\omega^{\prime\prime})\Big]
\end{eqnarray} 
Now using eq.(\ref{I2}) of appendix A we evaluate the frequency integral 
and obtain
\begin{eqnarray}
& &\langle\xi^{(0)}_j({\mathbf q},\omega)
\xi^{(1)}_k({\mathbf q}^\prime,\omega^\prime)\rangle\nonumber\\
&=&-\frac{\lambda\omega_h^2\gamma^2}{2ma^2W_2W_3}
D_k{(\mathbf q^{\prime},\omega^{\prime})}
\delta_{{\mathbf q}+{\mathbf q}^\prime,0}
\delta(\omega+\omega^{\prime})\nonumber\\
& &\times\sum_{l,j_1,j_2,j_3=1}^N\sum_{{\mathbf q}^{\prime\prime}}
z_l\nonumber\\
& &\times\Big[X_{kj_1j_2j_3}({\mathbf q}^\prime,{\mathbf q}^\prime,
{\mathbf q}^{\prime\prime})G^{(0)}_{jj_1}({\mathbf q},\omega)
\frac{A_{l,j_2}A_{l,j_3}}{B_{j_2j_3}({\mathbf q}^{\prime\prime})}\nonumber\\
&&+ X_{kj_1j_2j_3}({\mathbf q}^{\prime},{\mathbf q}^{\prime\prime},
{\mathbf q}^{\prime}) G^{(0)}_{jj_2}({\mathbf q},\omega)
\frac{A_{l,j_1}A_{l,j_3}}{B_{j_1j_3}({\mathbf q}^{\prime\prime})}\nonumber\\
&&+ X_{kj_1j_2j_3}({\mathbf q}^{\prime},{\mathbf q}^{\prime\prime},
-{\mathbf q}^{\prime\prime})G^{(0)}_{jj_3}({\mathbf q},\omega)
\frac{A_{l,j_1}A_{l,j_2}}{B_{j_1j_2}({\mathbf q}^{\prime\prime})}\Big]
\end{eqnarray} 
Then we use eq.(\ref{X}), (\ref{calM0}) and (\ref{calM1}) to obtain
\begin{equation}
\langle\xi^{(0)}_j({\mathbf q},\omega)
\xi^{(1)}_k({\mathbf q}^\prime,\omega^\prime)\rangle
= G^{(1)}_{jk}({\mathbf q}, \omega)
\delta_{{\mathbf q}+{\mathbf q}^\prime,0}
\delta(\omega+\omega^\prime),
\label{xi01w}
\end{equation}
where
\begin{eqnarray}
& &G^{(1)}_{jk}({\mathbf q},\omega)\nonumber\\
&=& -\frac{8\lambda\omega^2_h\gamma^2}{ma^2}
\frac{N}{(N+1)^3}D_k(-{\mathbf q},-\omega)
\sum_{l,j_1,j_2,j_3=1}^N \sum_{s,s_2,s_3=\pm}\nonumber\\
& &\times s s_2 s_3
\, z_l\sin(lj_2\nu)\sin(lj_3\nu)
\tilde{\delta}_{j_1+s_2j_2+s_3j_3+sk}
G^{(0)}_{jj_1}({\mathbf q},\omega)\nonumber\\
& &\times\Big\{{\cal M}^{(0)}_{j_2j_3}
\cos\frac{j_1\nu}{2}\cos\frac{j_2\nu}{2}
\cos\frac{j_3\nu}{2}\cos\frac{k\nu}{2}\nonumber\\
& &+(\sin^2\frac{q_2a}{2}+\sin^2\frac{q_3a}{2})
{\cal M}^{(1)}_{j_2j_3}\Big\},
\label{G1}
\end{eqnarray}
and $\tilde{\delta}_{j_1+s_2j_2+s_3j_3+sk}$
in terms of Kronecker delta functions reads
\begin{eqnarray}
& &\tilde{\delta}_{j_1+s_2j_2+s_3j_3+sk}\nonumber\\
&=&\delta_{j_1+s_2j_2+s_3j_3+sk,0}
+\delta_{j_2+s_2j_1+s_3j_3+sk,0}\nonumber\\
& &+\delta_{j_3+s_2j_2+s_3j_1+sk,0}.
\label{deltatilde}
\end{eqnarray}

\section{Radiated heat and temperature profile}

In the steady state limit, 
the mean square velocity of a particle in the layer 
at $n_1$ reads as
\begin{eqnarray}
v^2(n_1) &=& \frac{1}{W_2W_3}\sum_{{\mathbf n}_\perp}
\langle \dot{x}^2_{{\mathbf n}}\rangle\nonumber\\
&=&\frac{1}{W_2W_3}\sum_{{\mathbf q}}\sum_{k_1,k_2=1}^N
A_{n_1,k_1}A_{n_1,k_2}\nonumber\\
& &\times\langle\dot{\xi}_{k_1}({\mathbf q},t)
\dot{\xi}_{k_2}(-{\mathbf q},t)\rangle,
\end{eqnarray}
where we have used eq.(\ref{fexnt}) and 
(\ref{yxi}) in the last step. 
Then we use eq.(\ref{xit}), (\ref{xi00w}) and 
(\ref{xi01w}) and obtain the right hand side in the 
frequency space as
\begin{eqnarray}
v^2(n_1) &=& 
\frac{1}{2\pi W_2W_3}\sum_{\mathbf q}\sum_{k_1,k_2=1}^N 
A_{n_1,k_1}A_{n_1,k_2}
\int_{-\infty}^{\infty}\frac{d\omega}{2\pi}\nonumber\\
& &\times\omega^2[G^{(0)}_{k_1k_2}({\mathbf q},\omega)+
2G^{(1)}_{k_1k_2}({\mathbf p},\omega)+O(\lambda^2)].\nonumber\\
& &
\end{eqnarray}
We use eq.(\ref{xi00w}) and (\ref{xi01w}) and 
evaluate the frequency integrals using eq.(\ref{I4}) 
and (\ref{I6}). Then we use the results of appendix B and the 
definition of integrals in appendix C to write 
the final expression which reads in the continuum limit as
\begin{eqnarray}
v^2(n_1) &=& \frac{2}{m}\Big\{
\sum_{l=1}^N C^{(0)}_{n_1,l}z_l
+\frac{32\lambda}
{ma^2\omega^2_h}\sum_{l,l^\prime=1}^N
C^{(1)}_{n_1,l,l^\prime}z_l z_{l^\prime}\nonumber\\
& &+0(\lambda^2)\Big\}
\label{msvelocity}
\end{eqnarray}
where 
\begin{eqnarray}
C^{(0)}_{n_1,l}&=&\frac{1}{(N+1)^2}\sum_{k_1,k_2=1}^N
\sin(n_1k_1\nu)\sin(n_1k_2\nu)\nonumber\\
& &\times\sin(lk_1\nu)\sin(lk_2\nu)
\frac{\Lambda_{k_1k_2}}{\Delta_{k_1k_2}},\\
C^{(1)}_{n_1,l,l^\prime} &=&
\frac{N}{(N+1)^5}\sum_{k_1,k_2,j_1,j_2,j_3=1}^N
\sum_{s,s_2,s_3=\pm}s s_2 s_3\nonumber\\
& &\times\sin(n_1k_1\nu)\sin(n_1k_2\nu)
\sin(lj_2\nu)\sin(lj_3\nu)\nonumber\\
& &\times\sin(l^\prime k_1\nu)\sin(l^\prime j_1\nu)
\tilde{\delta}_{j_1+s_2j_2+s_3j_3+sk_2}\nonumber\\
& &\times V(k_1,k_2,j_1,j_2,j_3).
\end{eqnarray}
The expression for $\Delta_{k_1k_2}$ is given in eq.(\ref{Delta}).
The expressions for $\Lambda_{k_1k_2}$ and 
$V(k_1,k_2,j_1,j_2,j_3)$ are given in the following:
\begin{eqnarray}
\Lambda_{k_1k_2} &=& \Delta_{k_1k_2}
-(\cos k_1\nu-\cos k_2\nu)^2
F(1/2,1/2,1;\nonumber\\
& &(4\tilde{\gamma}^2/\Delta_{k_1k_2})^2)
\end{eqnarray}
and
\begin{eqnarray}
& &V(k_1,k_2,j_1,j_2,j_3) =\tilde{\gamma}^4\Big[4 g^{(2)}_{k_1k_2j_1}
\Big\{g^{(1)}_{j_1j_2j_3k_2}M^{(3)}_{k_1k_2j_1}
M^{(0)}_{j_2j_3}\nonumber\\
& &+\frac{1}{2}M^{(4)}_{k_1k_2j_1}M^{(1)}_{j_2j_3}\Big\}
+g^{(3)}_{k_1k_2j_1}\Big\{g^{(1)}_{j_1j_2j_3k_2}M^{(2)}_{k_1k_2j_1}
M^{(0)}_{j_2j_3}\nonumber\\
& &+\frac{1}{2}M^{(3)}_{k_1k_2j_1}M^{(1)}_{j_2j_3}\Big\}\Big]
\end{eqnarray}
where
\begin{eqnarray}
g^{(1)}_{j_1j_2j_3k_2} &=& \cos\frac{j_1\nu}{2} \cos\frac{j_2\nu}{2}
\cos\frac{j_3\nu}{2} \cos\frac{k_2\nu}{2},\\
g^{(2)}_{k_1k_2j_1} &=& 2\cos k_1\nu -\cos k_2\nu -\cos j_1\nu,\\
g^{(3)}_{k_1k_2j_1} &=& (2\cos k_1\nu - \cos k_2\nu-\cos j_1\nu)
(2+\cos k_1\nu)\nonumber\\
& &+\cos^2 k_1\nu -\cos k_2\nu\cos j_1\nu.  
\end{eqnarray}
The integrals $M^{(a)}_{jk}$ ($a=0,\,1$) and $M^{(b)}_{jkl}$ 
($b=2,\, 3,\, 4$) are given in the appendices B and C.
We list the following properties for 
$V(k_1,k_2,j_1,j_2,j_3)$, $C^{(0)}_{j,k}$ and 
$C^{(1)}_{j,k,l}$:
\begin{eqnarray}
& & V(k_1,k_2,j_1,j_2,j_3) = V(k_1,j_1,k_2,j_2,j_3),\\
& & C^{(0)}_{j,k} = C^{(0)}_{k,j},~~~
C^{(0)}_{j,k}=C^{(0)}_{N+1-j,N+1-k}\\
{\rm and}~~& & C^{(1)}_{j,k,l}
= C^{(1)}_{N+1-j,N+1-k,N+1-l}.
\end{eqnarray}
In the steady state particles in a layer come to equilibrium
owing to interactions with the adjacent layers and the attached
heat bath. If $T_{n_1}$ be the temperature of the layer at $n_1$,
then according to equipartition theorem particle's mean kinetic 
energy
\begin{equation}
\frac{1}{2}m v^2(n_1) = \frac{1}{2}k_B T_{n_1}.
\label{equipartition}
\end{equation}
With the use of eq.(\ref{msvelocity}), we obtain the following equations for
$z_l$s ($l=1, \cdots, N$):
\begin{eqnarray}
& &\sum_{l=1}^N C^{(0)}_{n_1,l}z_l
+\frac{32\lambda}
{ma^2\omega^2_h}\sum_{l,l^\prime=1}^N
C^{(1)}_{n_1,l,l^\prime}z_l z_{l^\prime}
+0(\lambda^2)\nonumber\\
&=&\frac{1}{2}k_B T_{n_1}.
\label{zTeqn1}
\end{eqnarray}
It is clear that the entire slab is an assembly of $N$ layers
which are in equilibria at different temperatures.
Apart from the temperatures $T_1$ and $T_N$ maintained 
at the two ends of the slab, temperatures $T_j$ ($2\le j\le N-1$)
of the intermediate layers and $z_l$ ($l=1,\cdots N$) are 
unknown quantities in the above equation. 
Moreover, $\tilde{\gamma}$ is also an unknown quantity
in addition to those $2N-2$ unknowns and it makes the 
number of unknowns $2N-1$. Since the number of 
equations involving these $2N-1$ unknowns are $N$,
we need to know some more information regarding the system
in order to find the solutions of all the unknowns. 

The rate of energy transfer from a particle at
${\mathbf n}$ to the adjacent heat bath reads\cite{Bonetto2004}
\begin{equation}
R_{{\mathbf n}} = m\gamma\dot{x}^2_{{\mathbf n}}
-\dot{x}_{{\mathbf n}}\eta_{{\mathbf n}}.
\end{equation}
Now consider the layer at $n_1$. The average loss of energy 
per unit time from any particle in the layer to the adjacent heat bath
\begin{eqnarray}
G(n_1) &=& \frac{1}{W_2W_3}\sum_{{\mathbf n}_\perp}
\langle R_{\mathbf n}\rangle\nonumber\\
&=&-\frac{1}{W_2W_3}\sum_{{\mathbf n}_\perp}
\langle\dot{x}_{{\mathbf n}}\eta_{\mathbf n}\rangle
+m\gamma v^2(n_1)\nonumber\\
&=& -\frac{1}{W_2W_3}\sum_{{\mathbf q}}\sum_{k=1}^N A_{n_1,k}
\{\langle\dot{\xi}^{(0)}_k({\mathbf q},t)f_{n_1}({\mathbf q},t)\rangle
\nonumber\\
& &+\langle\dot{\xi}^{(1)}_k({\mathbf q},t)f_{n_1}({\mathbf q},t)\rangle
+0(\lambda^2)\}+\gamma k_BT_{n_1}
\end{eqnarray}
where we have used eq.(\ref{fexnt}), (\ref{feetant}),
(\ref{yxi}), (\ref{xit}) and (\ref{equipartition})
in the last step. Then using eq.(\ref{xi0w}), 
and the equations (\ref{xi1w})-(\ref{xi00w})
we obtain
\begin{eqnarray}
& &G(n_1)\nonumber\\
&=&\gamma k_BT_{n_1}\nonumber\\
& &-\frac{i\gamma z_{n_1}}{W_2W_3}
\sum_{\mathbf q}\sum_{k=1}^N A^2_{n_1,k}
\int_{-\infty}^\infty\frac{d\omega}{2\pi}
\omega D_k({\mathbf q},\omega)\nonumber\\
& &+\frac{i\lambda\omega_h^2\gamma z_{n_1}}{8\pi a^2(W_2W_3)^2}
\sum_{{\mathbf q},{\mathbf q}^\prime}
\sum_{k,k_1,k_2,k_3=1}^N A_{n_1,k}\nonumber\\
& &\times\int_{-\infty}^\infty\frac{d\omega}{2\pi}
\frac{d\omega^\prime}{2\pi}
\omega D_k({\mathbf q},\omega)\nonumber\\
& &\times\Big\{
A_{n_1,k_1}X_{kk_1k_2k_3}({\mathbf q},{\mathbf q},{\mathbf q}^\prime)
D_{k_1}({\mathbf q},\omega)G^{(0)}_{k_2k_3}({\mathbf q}^\prime,\omega^\prime)
\nonumber\\
& &
+A_{n_1,k_2}X_{kk_1k_2k_3}({\mathbf q},{\mathbf q}^\prime,{\mathbf q})
D_{k_2}({\mathbf q},\omega)G^{(0)}_{k_1k_3}({\mathbf q}^\prime,\omega^\prime)
\nonumber\\
& &
+A_{n_1,k_3}X_{kk_1k_2k_3}({\mathbf q},{\mathbf q}^\prime,-{\mathbf q}^\prime)
D_{k_3}({\mathbf q},\omega)G^{(0)}_{k_1k_2}({\mathbf q}^\prime,\omega^\prime)
\Big\}.\nonumber\\
& &
\end{eqnarray}
The $\lambda$ dependent terms involve a particular type of
frequency integral which contains a factor of 
$\omega D_k({\mathbf q},\omega)D_l({\mathbf q},\omega)$
as its integrand and the value of that integral after 
performing the integration over $\omega$ is zero
according to eq.(\ref{I7}). Thus the $\lambda$ order
contribution to the radiated heat remains zero.
Then, after the use of 
eq.(\ref{I1}), we obtain the radiated energy per unit time from
any particle in the layer at $n_1$
\begin{equation}
G(n_1)= \gamma(k_BT_{n_1}-z_{n_1})+0(\lambda^2),
\end{equation} 
where the same expression is obtained in Ref.\cite{Shila2013} 
for harmonic crystal.
The radiation will take place from different layers 
of the slab. We assume that the Newton's law of cooling holds
for the radiated heat and set
\begin{equation}
z_{n_1}=k_B T_e~~{\rm for}~~2\le n_1\le N,
\label{znforngt1}
\end{equation}
where $T_e$ is the temperature of the environment in
which the slab is placed.
As a consequence of this, the number of unknowns in eq.(\ref{zTeqn1})
are now same as the number of equations.  
We use this result in eq.(\ref{zTeqn1}) and caste the 
equation in the following form in terms of dimensionless quantities.
\begin{equation}
H^{(2)}_{n_1}\bar{z}^2_1 +H^{(1)}_{n_1}\bar{z}_1
+H^{(0)}_{n_1}=\frac{1}{2}\bar{T}_{n_1},
\label{zTeqn2}
\end{equation}
where
\begin{eqnarray}
\bar{z}_1 &=& \frac{z_1}{k_BT_D},\nonumber\\
H^{(0)}_{n_1} &=& \bar{T}_e\sum_{l=2}^N C^{(0)}_{n_1,l}
+32\bar{\lambda}\bar{T}^2_e
\sum_{l,l^\prime =2}^N C^{(1)}_{n_1,l,l^\prime},\\
H^{(1)}_{n_1} &=& C^{(0)}_{n_1,1}
+32\bar{\lambda}\bar{T}_e
\sum_{l=2}^N(C^{(1)}_{n_1,1,l}+C^{(1)}_{n_1,l,1}),\\
H^{(2)}_{n_1} &=& 32\bar{\lambda}
C^{(1)}_{n_1,1,1}.
\end{eqnarray}
For a given material the Debye temperature
\begin{equation}
T_D=\frac{\hbar\omega_h}{k_B}~~
{\rm and}~~
\lambda_0=\frac{ma^2\omega^2_h}{k_BT_D}
=\frac{ma^2k_BT_D}{\hbar^2}.
\end{equation} 
We have defined
\begin{equation}
\bar{\lambda}=\frac{\lambda}{\lambda_0},~~
\bar{T}_e=\frac{T_e}{T_D}
~~{\rm and}~~ 
\bar{T}_{n_1}=\frac{T_{n_1}}{T_D}.
\end{equation}
We subtract the equation for $n_1=N$ from the equation for
$n_1=1$ of eq.(\ref{zTeqn2}) and obtain $\bar{z}_1$ to 
order $\bar{\lambda}$ as
\begin{eqnarray}
\bar{z}_1 &=& \bar{T}_e+\frac{\bar{T}_1-\bar{T}_N}
{2(C^{(0)}_{1,1}-C^{(0)}_{N,1})}\Big[1
-\frac{32\bar{\lambda}}
{C^{(0)}_{1,1}-C^{(0)}_{N,1}}\nonumber\\
& &\times\Big\{\frac{C^{(1)}_{1,1,1}-C^{(1)}_{N,1,1}}
{2(C^{(0)}_{1,1}-C^{(0)}_{N,1})}
(\bar{T}_1-\bar{T}_N)\nonumber\\
& &+\bar{T}_e\sum_{j=1}^{N-1}(C^{(1)}_{1,j,1}+C^{(1)}_{1.1,j}
-C^{(1)}_{N,j,1}-C^{(1)}_{N,1,j})\Big\}\nonumber\\
& &+0(\bar{\lambda}^2)\Big].
\label{z1bar}
\end{eqnarray}
We see from eq.(\ref{zTeqn2}) that the following relation that 
\begin{eqnarray}
& &\frac{\bar{T}_j-\bar{T}_{N+1-j}}{\bar{T}_1-\bar{T}_N}\nonumber\\
&=&r_1\frac{C^{(0)}_{j,1}-C^{(0)}_{N+1-j,1}}{C^{(0)}_{1,1}-C^{(0)}_{N,1}}
+r_2\frac{C^{(1)}_{j,1,1}-C^{(1)}_{N+1-j,1,1}}{C^{(0)}_{1,1}-C^{(0)}_{N,1}}
\nonumber\\
& &+r_3\frac{\sum_{l=1}^{N-1}(C^{(1)}_{j,1,l}+C^{(1)}_{j,l,1} 
-C^{(1)}_{N+1-j,1,l}-C^{(1)}_{N+1-j,l,1})}{C^{(0)}_{1,1}-C^{(0)}_{N,1}}.
\nonumber\\
& &
\label{Trelation}
\end{eqnarray}
holds. The expressions for $r_1$, $r_2$ and $r_3$ are given as
\begin{eqnarray}
r_1 &=& 1-32\bar{\lambda}\Big\{
\frac{C^{(1)}_{1,1,1}-C^{(1)}_{N,1,1}}{2(C^{(0)}_{1,1}-C^{(0)}_{N,1})^2}
(\bar{T}_1-\bar{T}_N)\nonumber\\
& &+\bar{T}_e\frac{\sum_{l=1}^{N-1}(C^{(1)}_{1,1,l}+C^{(1)}_{1,l,1} 
-C^{(1)}_{N,1,l}-C^{(1)}_{N,l,1})}{C^{(0)}_{1,1}-C^{(0)}_{N,1}}\Big\}
\nonumber\\
& &+0(\bar{\lambda}^2),\\
r_2 &=& 16\bar{\lambda}
\frac{\bar{T}_1-\bar{T}_N}{C^{(0)}_{1,1}-C^{(0)}_{N,1}}
+0(\bar{\lambda}^2),\\
r_3 &=& 32\bar{\lambda}\bar{T}_e
+ 0(\bar{\lambda}^2) .
\end{eqnarray}
We obtain the required temperature profile consistent with this 
relation in eq.(\ref{Trelation}) as 
\begin{eqnarray}
\bar{T}_j &=& \frac{\bar{T}_1+\bar{T}_N}{2}
+\frac{\bar{T}_1-\bar{T}_N}{C^{(0)}_{1,1}-C^{(0)}_{N,1}}
\Big\{r_1\Big(C^{(0)}_{j,1}-\frac{C^{(0)}_{1,1}+C^{(0)}_{N,1}}{2}\Big)
\nonumber\\
& &+r_2\Big(C^{(1)}_{j,1,1}-\frac{C^{(1)}_{1,1,1}+C^{(1)}_{N,1,1}}{2}\Big)
\nonumber\\
& &+r_3\sum_{l=1}^{N-1}\Big(C^{(1)}_{j,1,l}+C^{(1)}_{j,l,1}\nonumber\\
& &-\frac{C^{(1)}_{1,1,l}+C^{(1)}_{1,l,1}+C^{(1)}_{N,1,l}+C^{(1)}_{N,l,1}}{2}
\Big)\Big\}.
\end{eqnarray}
It has been observed\cite{Shila2013} that $C^{(0)}_{j,1}=\alpha^\prime_0
+\alpha_0 e^{-b_0|j-1|}$, where 
$\alpha_0 \rightarrow 4.99 \times 10^{-3}$
and $ b_0\rightarrow 0.031$ for $\gamma=0.01$ 
in the thermodynamic limit of a three dimensional system.
It is found that $\alpha^\prime_0$, though much smaller than 
$\alpha_0$, is $N$ dependent. We use this form 
of $C^{(0)}_{j,1}$ in the above equation, and obtain 
the temperature profile as
\begin{eqnarray}
\bar{T}_j &=& \bar{T}_N + (\bar{T}_1-\bar{T}_N)
\Big\{r_1 e^{-b_0|j-1|}
+\frac{r_2\chi^{(1)}_j+r_3\chi^{(2)}_j}
{C^{(0)}_{1,1}-C^{(0)}_{N,1}}\Big\},\nonumber\\
& &
\end{eqnarray}
where
\begin{eqnarray}
\chi^{(1)}_j &=& C^{(1)}_{j,1,1}-C^{(1)}_{N,1,1}\\
{\rm and}~~\chi^{(2)}_j &=& \sum_{l=1}^{N-1}(C^{(1)}_{j,1,l}+C^{(1)}_{j,l,1}
-C^{(1)}_{N,1,l}-C^{(1)}_{N,l,1}).
\end{eqnarray}
The leading behaviour of the temperature profile has an exponentially 
falling nature from high to low temperature end of the slab and 
this nature will be modified by the functions $\chi^{(1)}_j$ and
$\chi^{(2)}_j$ in the non-leading order. We note that the 
$\bar{\lambda}$ dependent terms of the temperature
profile are either proportional to $(T_1-T_N)^2$ 
or proportional to $T_e(T_1-T_N)$.

We evaluate numerically $\chi^{(a)}_j$s ($a=1,2$) as a function of 
$j$ using our available resources for $N=16$ and $N=20$ and 
for $\gamma=0.01$. Then we fit the curves with the known functions.
The obtained data points together with their fitted curves 
are plotted in Fig.1, 2, 3 and 4. 
\begin{figure}[!th]
\begin{center}
\includegraphics*[width=0.48 \textwidth,clip=true]{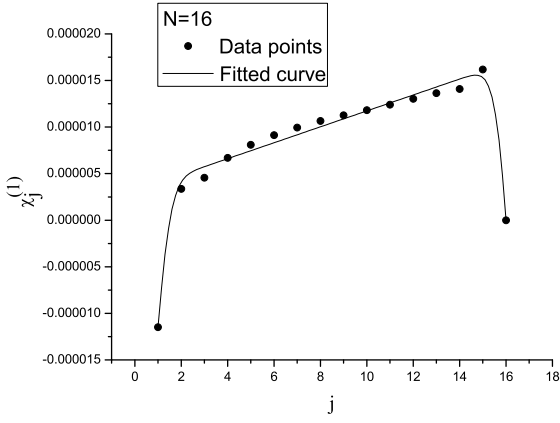}
\end{center}
\vskip -0.4in
\caption[]{Plot of data points and the fitted curve of
$\chi^{(1)}_j$ vs $j$ for $N=16$ and $\gamma=0.01$.}
\label{plotchi1j16}
\end{figure}
\begin{figure}[!th]
\begin{center}
\includegraphics*[width=0.48 \textwidth,clip=true]{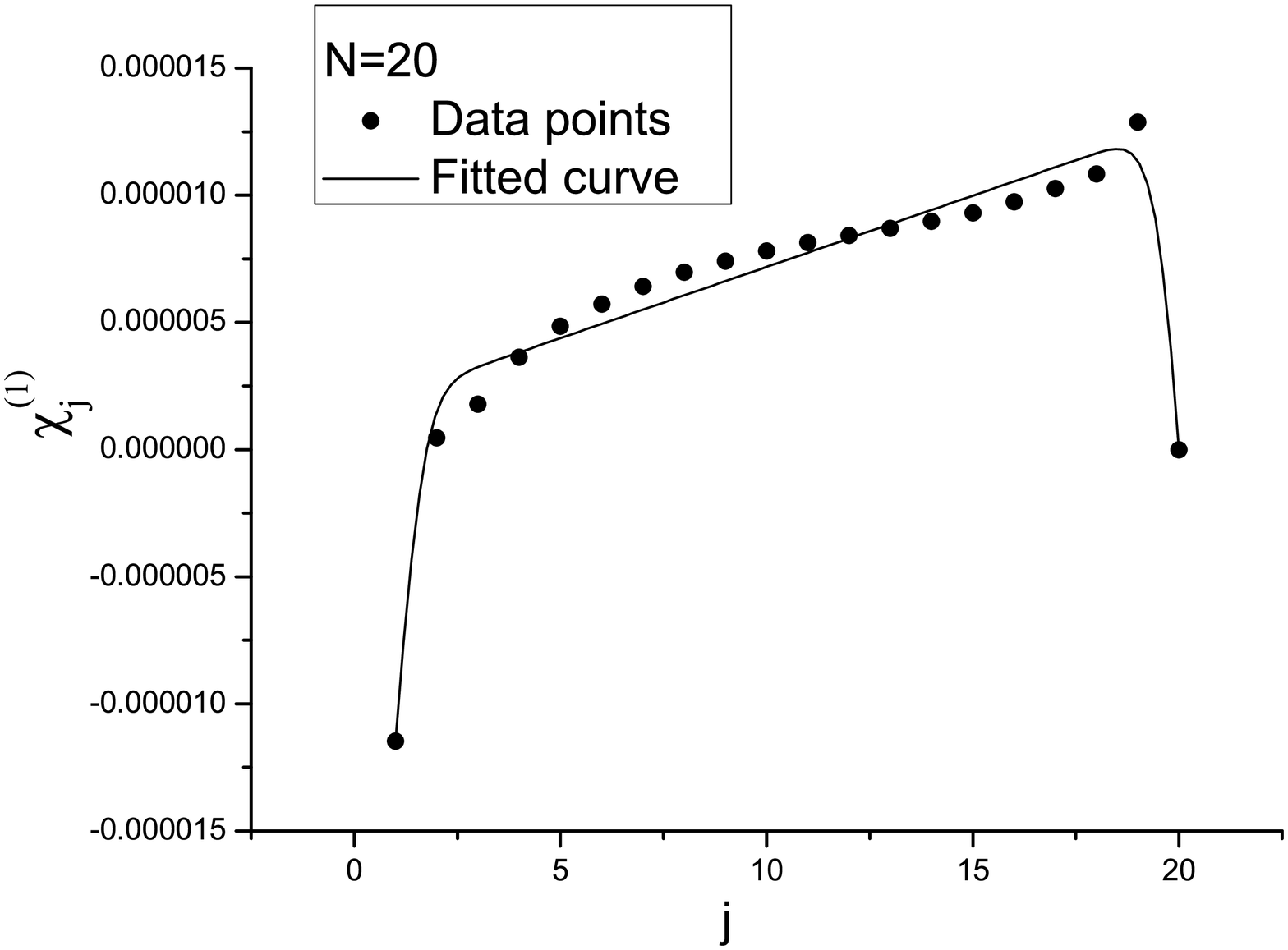}
\end{center}
\vskip -0.4in
\caption[]{Plot of data points and the fitted curve of
$\chi^{(1)}_j$ vs $j$ for $N=20$ and $\gamma=0.01$.}
\label{plotchi1j20}
\end{figure}
\begin{figure}[!th]
\begin{center}
\includegraphics*[width=0.48 \textwidth,clip=true]{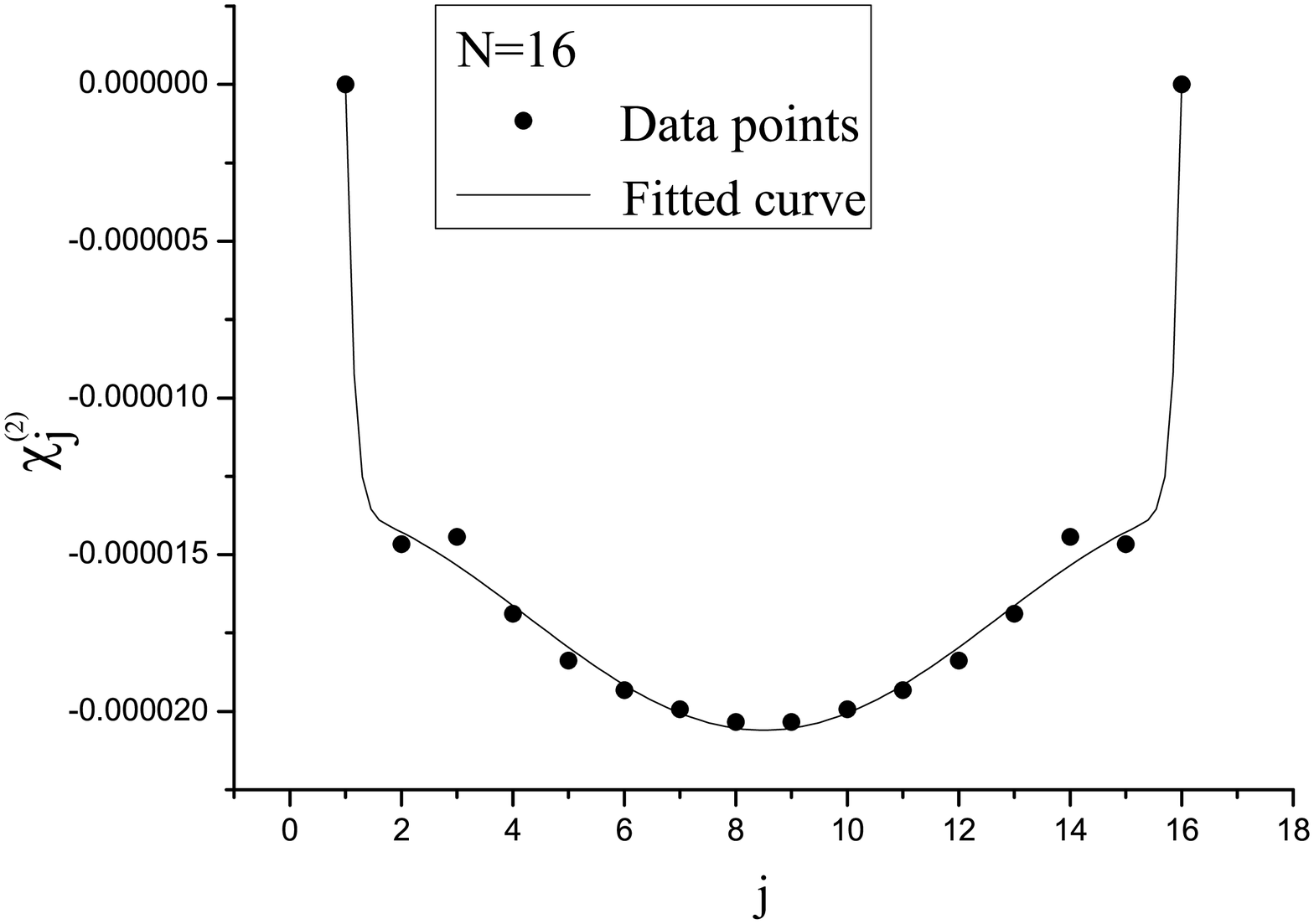}
\end{center}
\vskip -0.4in
\caption[]{Plot of data points and the fitted curve of
$\chi^{(2)}_j$ vs $j$ for $N=16$ and $\gamma=0.01$.}
\label{plotchi2j16}
\end{figure}
\begin{figure}[!th]
\begin{center}
\includegraphics*[width=0.48 \textwidth,clip=true]{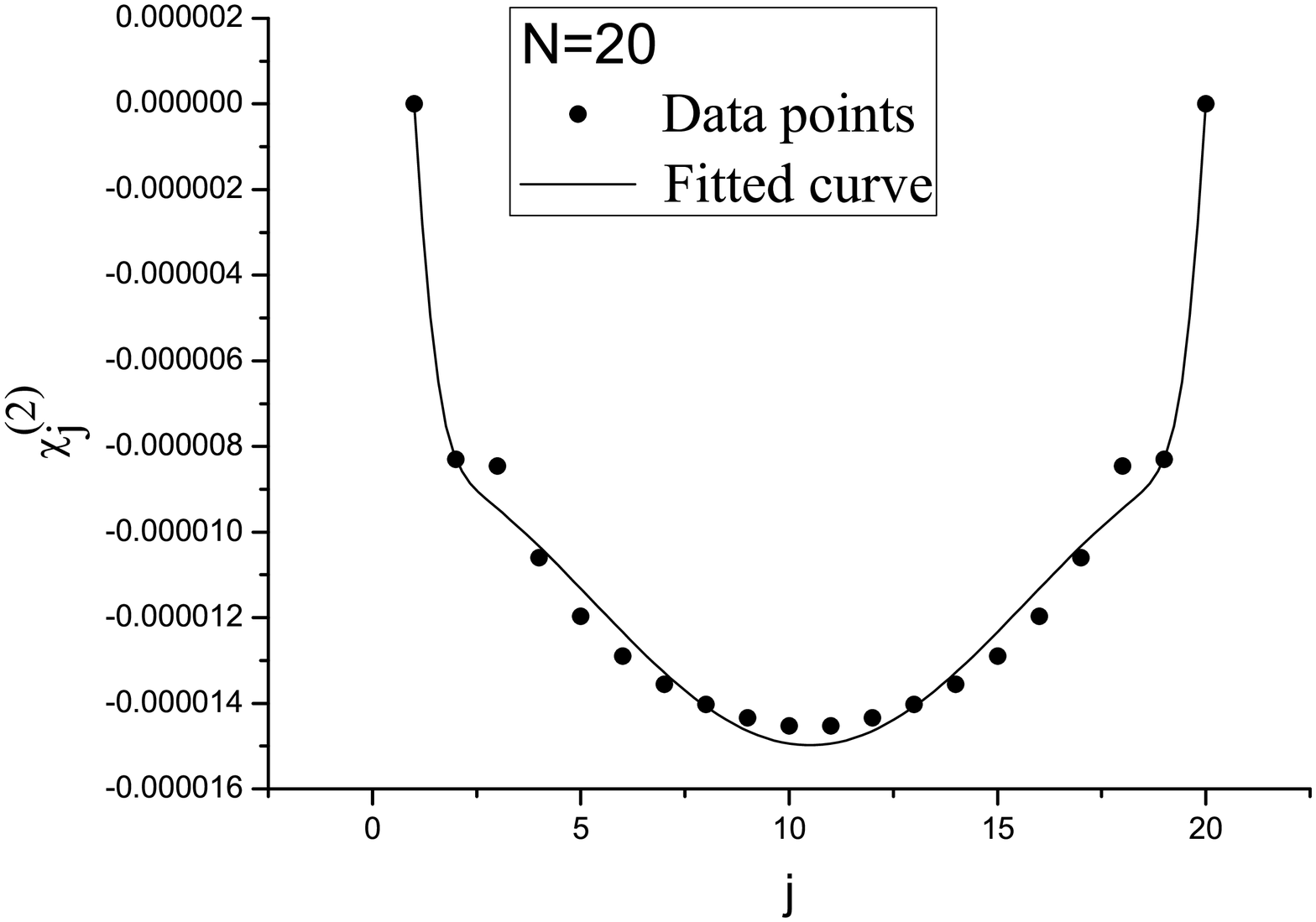}
\end{center}
\vskip -0.4in
\caption[]{Plot of data points and the fitted curve of
$\chi^{(2)}_j$ vs $j$ for $N=20$ and $\gamma=0.01$.}
\label{plotchi2j20}
\end{figure}
The obtained functional forms of 
$\chi^{(a)}_j$s ($a=1,2$) are as follows:
\begin{eqnarray}
\chi^{(1)}_j &=&\frac{1+\eta_1j}{1+\eta_1}
\Big\{\alpha_1(e^{-b_1}+e^{-b_1N^2} 
-e^{-b_1j^2}\nonumber\\
& &-e^{-b_1(N+1-j)^2})
+\chi^{(1)}_1\frac{e^{-b_1j^2}-e^{-b_1N^2}}{e^{-b_1}-e^{-b_1N^2}}
\Big\},\\
\chi^{(2)}_j &=& \alpha_2
\frac{1+\eta_2\sin^2j\nu}{1+\eta_2\sin^2\nu}
\Big\{e^{-b_2j^2}+e^{-b_2(N+1-j)^2} 
-e^{-b_2}\nonumber\\
& &-e^{-b_2N^2}\Big\}, 
\end{eqnarray}
where the property $\chi^{(2)}_j=\chi^{(2)}_{N+1-j}$ holds.
We have shown in the next section that $\chi^{(1)}_1$ tends
to a finite value in the thermodynamic limit.
The value of the parameters for two different $N$ are given
in the following two tables.
\begin{table}[!th]
{\begin{tabular}{|c|c|c|c|}
\hline
$N$ & $\alpha_1$ & $b_1$ & $\eta_1$\\
\hline
$16$ & $11.48\times 10^{-6}$ & $1.05$ & $0.27$\\
\hline
$20$ & $5.16\times 10^{-6}$ & $0.88$ & $0.35$\\
\hline
\end{tabular}}
\caption{Parameters of $\chi^{(1)}_j$ for different $N$
and $\gamma=0.01$}
\label{tablechi1j}
\end{table}
\begin{table}[!th]
{\begin{tabular}{|c|c|c|c|}
\hline
$N$ & $\alpha_2$ & $b_2$ & $\eta_2$\\
\hline
$16$ & $43.53\times 10^{-5}$ & $3.47$ & $0.55$\\
\hline
$20$ & $2.18\times 10^{-5}$ & $0.96$ & $0.83$\\
\hline
\end{tabular}}
\caption{Parameters of $\chi^{(2)}_j$ for different $N$
and $\gamma=0.01$}
\label{tablechi2j}
\end{table}

\section{Conduction current and Fourier's law}

The energy current density flown between ${\mathbf n}$ and
${\mathbf n}+\hat{e}_1$, where $\hat{e}_1=(1,0,0)$, reads
as
\begin{equation}
j_{\mathbf n}=\frac{1}{2 a^2}(\dot{x}_{{\mathbf n}+\hat{e}_1} 
+\dot{x}_{\mathbf n})\{C_2(x_{\mathbf n}-x_{{\mathbf n}+\hat{e}_1})
+C_4(x_{\mathbf n}-x_{{\mathbf n}+\hat{e}_1})^3\}.
\end{equation}
The average energy current density per bond along $n_1$ direction 
\begin{equation}
J=\frac{1}{W_2W_3(N-1)}\sum_{{\mathbf n}_\perp}\sum_{n_1=1}^N
j_{\mathbf n} = J_1+J_2,
\end{equation}
where 
\begin{eqnarray}
J_1 &=&\frac{C_2}{2 a^2W_2W_3(N-1)}\sum_{{\mathbf n}_\perp}\sum_{n_1=1}^N
(\dot{x}_{{\mathbf n}+\hat{e}_1}+\dot{x}_{\mathbf n})\nonumber\\
& &\times (x_{\mathbf n}-x_{{\mathbf n}+\hat{e}_1}),
\label{J1def}\\
J_2 &=& \frac{C_4}{2 a^2W_2W_3(N-1)}\sum_{{\mathbf n}_\perp}\sum_{n_1=1}^N
(\dot{x}_{{\mathbf n}+\hat{e}_1}+\dot{x}_{\mathbf n})\nonumber\\
& &\times (x_{\mathbf n}-x_{{\mathbf n}+\hat{e}_1})^3.
\label{J2def}
\end{eqnarray}

\subsection{Calculation of $J_1$ }

We use the Fourier expansion of the the displacement fields from 
eq.(\ref{fexnt}) and also use eq.(\ref{yxi}), (\ref{xit}), 
(\ref{xi00w}), (\ref{xi01w}) and (\ref{dsum1}) to obtain the 
expression of $J_1$ in eq.(\ref{J1def}) to order $\lambda$ as
\begin{eqnarray}
J_1 &=& \frac{C_2}{2\pi a^2W_2W_3(N^2-1)}\sum_{\mathbf q}
\sum_{k_1,k_2=1}^N (1-(-1)^{k_1+k_2})\nonumber\\
& &\times\sin k_1\nu\sin k_2\nu
\int_{-\infty}^\infty\frac{d\omega}{2\pi}i\omega\nonumber\\
& &\times\Big[\Big(1-\frac{1}{\cos k_1\nu-\cos k_2\nu}\Big)
G^{(0)}_{k_1k_2}({\mathbf q},\omega)\nonumber\\
& &-\frac{2}{\cos k_1\nu-\cos k_2\nu}
G^{(1)}_{k_1k_2}({\mathbf q},\omega)
+0(\lambda^2)\Big].
\end{eqnarray}
We first use eq.(\ref{G0}) and (\ref{G1}) and evaluate 
the frequency integrals using the results in eq.(\ref{I3}) 
and (\ref{I5}). Then using the results of appendix B 
we obtain the expression as
\begin{eqnarray}
J_1 &=& \frac{\gamma k_BT_D}{(N-1)a^2}\Big[
\sum_{l=1}^N K^{(0)}_l\bar{z}_l
+32\bar{\lambda}
\sum_{l,l^\prime=1}^N K^{(1)}_{l,l^\prime}
\bar{z}_l\bar{z}_{l^\prime}\nonumber\\
& &+ 0(\bar{\lambda}^2)\Big],
\label{J1final}
\end{eqnarray}
where the harmonic part\cite{Shila2013}
\begin{eqnarray}
K^{(0)}_l &=& \frac{1}{(N+1)^2}\sum_{j,k=1}^N (1-(-1)^{j+k})
\sin j\nu\sin k\nu\nonumber\\
& &\times\sin(lj\nu)\sin(lk\nu) M^{(0)}_{jk}
\label{K0}
\end{eqnarray}
and the anharmonic part
\begin{eqnarray}
K^{(1)}_{l,l^\prime} &=& \frac{N}{(N+1)^5}\sum_{k_1,k_2,j_1,j_2,j_3=1}^N
\sum_{s,s_2,s_3=\pm}s s_2 s_3\nonumber\\
& &\times\sin k_1\nu\sin k_2\nu\sin(lj_2\nu)\sin(lj_3\nu)\nonumber\\
& &\times\sin(l^\prime j_1\nu)\sin(l^\prime k_1\nu)
\tilde{\delta}_{j_1+s_2j_2+s_3j_3+sk_2}\nonumber\\
& &\times U(k_1,k_2,j_1,j_2,j_3).
\label{K1}
\end{eqnarray}
The expression for $U(k_1,k_2,j_1,j_2,j_3)$ reads as
\begin{eqnarray}
& &U(k_1,k_2,j_1,j_2,j_3) =
\tilde{\gamma}^2(1-(-1)^{k_1+k_2})
\Big[(g^{(2)}_{k_1k_2j_1}-\tilde{\gamma}^2)\nonumber\\
& &\times\Big\{g^{(1)}_{j_1j_2j_3k_2}
M^{(2)}_{k_1k_2j_1}M^{(0)}_{j_2j_3}
+\frac{1}{2}M^{(3)}_{k_1k_2j_1}M^{(1)}_{j_2j_3}\Big\}\nonumber\\
& &-\frac{1}{\cos k_1\nu-\cos k_2\nu}
\Big\{g^{(1)}_{j_1j_2j_3k_2} M^{(0)}_{k_1j_1} M^{(0)}_{j_2j_3}\nonumber\\
& &+\frac{1}{2} M^{(1)}_{k_1j_1} M^{(1)}_{j_2j_3}\Big\}\Big]
\end{eqnarray}
Although there is a factor of $\cos k_1\nu-\cos k_2\nu$ in the 
denominator of the second term of $U(k_1,k_2,j_1,j_2,j_3)$, 
owing to the following result that 
\begin{equation}
\lim_{k_1\rightarrow k_2}
\frac{(1-(-1)^{k_1+k_2})}{\cos k_1\nu-\cos k_2\nu}
=0,
\end{equation}
it remains finite and goes to zero 
in the limit when $k_1$ goes to $k_2$.

\subsection{Calculation of $J_2$}

We obtain the expression of $J_2$ to order $\lambda$ after
using eq.(\ref{fexnt}), (\ref{yxi}), (\ref{xit}), 
(\ref{xi00w}) and (\ref{dsum2}) in eq.(\ref{J2def}) as
\begin{eqnarray}
J_2 &=& \frac{3m\omega_h^2\lambda}{32\pi^2 a^4(W_2W_3)^2(N-1)}
\sum_{k_1,k_2,k_3,k_4=1}^N\nonumber\\
& &\times{\cal N}(k_1,k_2,k_3,k_4)
\sum_{{\mathbf q},{\mathbf q}^\prime}
\int_{-\infty}^\infty\frac{d\omega}{2\pi}
\frac{d\omega^\prime}{2\pi} 
i\omega G^{(0)}_{k_1k_2}({\mathbf q},\omega)\nonumber\\
& &\times G^{(0)}_{k_3k_4}({\mathbf q}^\prime,\omega^\prime)
+0(\lambda^2),
\end{eqnarray}
where we have used the symmetry properties of ${\cal N}(k_1,k_2,k_3,k_4)$
given in eqn.(\ref{propcalN}). We carry out the frequency integrations
using eq.(\ref{I2}) and (\ref{I3}) and then use the results of appendix B 
to arrive at the expression 
\begin{equation}
J_2=\frac{48\bar{\lambda}\gamma k_BT_D}
{(N-1)a^2}\sum_{l,l^\prime=1}^N K^{(2)}_{l,l^\prime}
\bar{z}_l\bar{z}_{l^\prime} + 0(\bar{\lambda}^2),
\label{J2final}
\end{equation}
where
\begin{eqnarray}
K^{(2)}_{l,l^\prime} &=& -\frac{\tilde{\gamma}^2}{(N+1)^4}
\sum_{k_1,k_2,k_3,k_4=1}^N
\{1-(-1)^{k_1+k_2+k_3+k_4}\}\nonumber\\
& &\times{\cal N}_1(k_1,k_2,k_3,k_4)(\cos k_1\nu-\cos k_2\nu)
M^{(0)}_{k_1k_2}\nonumber\\
& &\times M^{(0)}_{k_3k_4}\sin(lk_1\nu)\sin(lk_2\nu)
\sin(l^\prime k_3\nu)\sin(l^\prime k_4\nu).\nonumber\\
& &
\label{K2}
\end{eqnarray}
The expression for ${\cal N}_1(k_1,k_2,k_3,k_4)$ is given in 
eq.(\ref{calN1}).

\subsection{Thermal conductivity}

The final expression according to eq.(\ref{J1final}) 
and (\ref{J2final}) for average energy current density per bond 
reads as
\begin{eqnarray}
J &=& \frac{\gamma k_BT_D}{(N-1)a^2}\Big[
\sum_{l=1}^N K^{(0)}_l\bar{z}_l +16\bar{\lambda}
\sum_{l,l^\prime=1}^N (2\,K^{(1)}_{l,l^\prime}\nonumber\\
& & 3\,K^{(2)}_{l,l^\prime})
\bar{z}_l\bar{z}_{l^\prime}
+ 0(\bar{\lambda}^2)\Big].
\label{Jfinal}
\end{eqnarray}
It is evident from the expressions given in eq.(\ref{K0}),
(\ref{K1}) and (\ref{K2}) for $K^{(0)}$, $K^{(1)}_{l,l^\prime}$
and $K^{(2)}_{l,l^\prime}$ respectively that the following properties 
hold for them.
\begin{eqnarray}
K^{(0)}_l &=& -K^{(0)}_{N+1-l},\\
K^{(1)}_{l,l^\prime} &=& -K^{(1)}_{N+1-l,N+1-l^\prime},\\
K^{(2)}_{l,l^\prime} &=& -K^{(1)}_{N+1-l,N+1-l^\prime}.
\end{eqnarray}
We then use these properties and eq.(\ref{znforngt1}) in
eq.(\ref{Jfinal}) to obtain 
\begin{eqnarray}
J &=& \frac{\gamma k_BT_D}{(N-1)a^2}
(\bar{z}_1-\bar{T}_e)\Big[K^{(0)}_1 +16\bar{\lambda}
\Big\{\bar{T}_e\sum_{l=2}^{N/2}\{2(K^{(1)}_{1,l}\nonumber\\
& &+K^{(1)}_{l,1} -K^{(1)}_{N,l} -K^{(1)}_{l,N})
+3(K^{(2)}_{1,l} +K^{(2)}_{l,1} -K^{(2)}_{N,l}\nonumber\\
& &-K^{(2)}_{l,N})\}
+(\bar{z}_1+\bar{T}_e)(2\,K^{(1)}_{1,1}+3\,K^{(2)}_{1,1})\Big\} 
+ 0(\bar{\lambda}^2)\Big],\nonumber\\
& &
\end{eqnarray}
where we have assumed $N$ as even.
Then after using eq.(\ref{z1bar}) we obtain the average
energy current density per bond as
\begin{equation}
J=\kappa\frac{T_1-T_N}{(N-1)a},
\end{equation}
where the thermal conductivity to order $\bar{\lambda}$
reads as
\begin{eqnarray}
\kappa &=& \frac{\gamma k_B}{2 a (C^{(0)}_{1,1}-C^{(0)}_{N,1})}
\Big[K^{(0)}_1\nonumber\\
& &+16\bar{\lambda}\Big\{\bar{T}_e\Big(\chi^{(4)}
- \frac{2 K_1^{(0)}\chi^{(2)}_1}{C^{(0)}_{1,1}-C^{(0)}_{N,1}}\Big)\nonumber\\
& &+\frac{\bar{T}_1-\bar{T}_N}{2(C^{(0)}_{1,1}-C^{(0)}_{N,1})}
\Big(\chi^{(3)}
- \frac{2 K_1^{(0)}\chi^{(1)}_1}{C^{(0)}_{1,1}-C^{(0)}_{N,1}}\Big)
\Big\}
+0(\bar{\lambda}^2)\Big].\nonumber\\
& &
\label{kappa}
\end{eqnarray}
The expressions for $\chi^{(3)}$ and $\chi^{(4)}$ are given in the following:
\begin{eqnarray}
\chi^{(3)} &=& 2K^{(1)}_{1,1}+3K^{(2)}_{1,1} \\
{\rm and}~~\chi^{(4)} &=& 
\sum_{l=1}^{N-1}\{2(K^{(1)}_{1,l} +K^{(1)}_{l,1})+
3(K^{(2)}_{1,l}+K^{(2)}_{l,1})\}.\nonumber\\
& &
\end{eqnarray}
We have evaluated $\chi^{(1)}_1$, $\chi^{(2)}_1$, $\chi^{(3)}$,
$\chi^{(4)}$ and $K^{(0)}_1$ numerically for different $N$ and 
taking $\gamma=0.01$.
Our evaluations give $\chi^{(2)}=0.0$ and $\chi^{(4)}=0.0$ for all $N$ 
and $K^{(0)}\rightarrow 18.09$ in the thermodynamic limit. 
The remaining results are given in the following table.
\begin{table}[!th]
{\begin{tabular}{|c|c|c|}
\hline
$N$ & $\chi^{(1)}_1$ & $\chi^{(3)}$\\
\hline
$10$ & $-1.08\times 10^{-5}$ & $-3.25\times 10^{-2}$ \\
\hline
$16$ & $-1.15\times 10^{-5}$ & $-3.63\times 10^{-2}$ \\
\hline
$20$ & $-1.15\times 10^{-5}$ & $-3.66\times 10^{-2}$ \\
\hline
$24$ & $-1.13\times 10^{-5}$ & $-3.6\times 10^{-2}$ \\
\hline
$30$ & $-1.09\times 10^{-5}$ & $-3.41\times 10^{-2}$ \\
\hline
$34$ & $-1.05\times 10^{-5}$ & $-3.26\times 10^{-2}$ \\     
\hline
$40$ & $-1.01\times 10^{-5}$ & $-3.01\times 10^{-2}$ \\
\hline
\end{tabular}}
\caption{$\chi^{(1)}_1$ and $\chi^{(3)}_1$ for different $N$
and $\gamma=0.01$}
\label{tablechi123}
\end{table}
It is clear that owing to vanishing values of $\chi^{(2)}_1$ and 
$\chi^{(4)}$ for any $N$, $\kappa$ remains insensitive to the 
environment temperature $T_e$ to order $\bar{\lambda}$. 
Since $\chi^{(1)}_1\rightarrow 10^{-5}$,
and $\chi^{(3)}\rightarrow 3\times 10^{-2}$ in the thermodynamic limit,
$\kappa$ receives a finite, non zero and
temperature dependent contribution which is linearly proportional 
to $T_1-T_N$ at its $\bar{\lambda}$ order. 
Thus $\kappa$ remains finite in the thermodynamic limit
and hence Fourier's law holds to order $\bar{\lambda}$.

\section{Summary}

We have taken an ordered, anharmonic, three dimensional crystal in the 
form of a slab geometry. We have attached Langevin type baths to the 
first and the $N$-th surfaces along its length maintaining the 
fixed temperatures $T_1$ and $T_N$ respectively. In order to simulate 
the environment surrounding the crystal, we have attached
heat baths of same type to the remaining $N-2$ surfaces of the slab.
We have chosen the noise functions of $N$ baths Gaussian and 
their widths $z_j$ ($1\le j\le N$) as site dependent.  
In the steady state, when $t\gg \frac{1}{\gamma}$, each surface behaves
as a thermally equilibriated system and the slab as a whole behaves 
as an assembly of $N$ such systems at different
temperatures. Our evaluation have shown that the average 
radiated energy per unit 
time from a particle in any one of the $N$ surfaces does not receive
any correction at the order $\bar{\lambda}$. We have taken $z_j$
as proportional to the environment temperature $T_e$ for $2\le j\le N$  
and have found that the radiated energies from $N-1$ layers obey Newton's
law of cooling. 
We have shown that the exponentially falling nature from
high to low temperature end of the slab is the leading 
behaviour of the temperature profile. The non-leading 
behaviour of the profile at the order $\bar{\lambda}$ is 
governed by the two functions $\chi^{(1)}_j$ and $\chi^{(2)}_j$.
Our numerical evaluations have shown that the 
thermal conductivity $\kappa$ remains independent of the 
environment temperature $T_e$ but it is dependent linearly on $T_1-T_N$ 
at the order $\bar{\lambda}$. Moreover, since $\kappa$ remains 
finite in the thermodynamic limit, Fourier's law holds to 
order $\bar{\lambda}$.

\appendix

\section{Frequency Integrals}

Consider the integral
\begin{eqnarray}
{\cal I}^{(1)}_j({\mathbf q})&=& \int^{\infty}_{-\infty}\frac{d\omega}{2\pi}
\omega D_j({\mathbf q},\omega)\nonumber\\
&=& -\int^\infty_{-\infty} \frac{d\omega}{2\pi} \frac{\omega}
{\omega^2-\omega^2_j({\mathbf q})-i\gamma\omega}. 
\end{eqnarray}
To evaluate this integral we consider the complex $\omega$ plane
and choose a  closed, semi circular contour which is closed in the 
upper half of the complex plane. Since the contour encloses the 
poles at $\pm\sqrt{\omega^2_k({\mathbf q})-\frac{\gamma^2}{4}}
+i\frac{\gamma}{2}$ the result of our evaluation is    
\begin{equation}
{\cal I}^{(1)}_j({\mathbf q}) = -i
\label{I1}
\end{equation}
Integrals of the following types are evaluated 
in the similar manner and results are quoted below.
\begin{equation}
{\cal I}^{(2)}_{jk}({\mathbf q})=\int^{\infty}_{\infty}\frac{d\omega}{2\pi}
D_j({\mathbf q},\omega)D_k(-{\mathbf q},-\omega)=\frac{2\gamma}
{B_{jk}(\mathbf q)},
\label{I2} 
\end{equation}
where
\begin{equation}
B_{jk}({\mathbf q}) = (\omega^2_j({\mathbf q})
-\omega^2_k({\mathbf q}))^2+
2\gamma^2 (\omega^2_j({\mathbf q})+\omega^2_k({\mathbf q})).
\label{Bjkq}
\end{equation}
\begin{eqnarray}
{\cal I}^{(3)}_{jk}({\mathbf q})&=&\int^\infty_{-\infty} \frac{d\omega}{2\pi}
 \omega D_j({\mathbf q},\omega)D_k(-{\mathbf q},-\omega)\nonumber\\
&=& \frac{i (\omega^2_j({\mathbf q})-\omega^2_k({\mathbf q}))}
{B_{jk}(\mathbf q)}.
\label{I3}
\end{eqnarray}
\begin{eqnarray}
{\cal I}^{(4)}_{jk}({\mathbf q})&=&\int^\infty_{-\infty} \frac{d\omega}{2\pi}
 \omega^2 D_j({\mathbf q},\omega)D_k(-{\mathbf q},-\omega)\nonumber\\
&=& \frac{\gamma(\omega^2_j({\mathbf q})+\omega^2_k({\mathbf q}))}
{B_{jk}(\mathbf q)}. 
\label{I4}
\end{eqnarray}
\begin{eqnarray}
{\cal I}^{(5)}_{jkl}({\mathbf q})
&=&\int^\infty_{-\infty} \frac{d\omega}{2\pi}
\omega D_j({\mathbf q},\omega)D_k(-{\mathbf q},-\omega)
D_l(-{\mathbf q},-\omega) \nonumber\\
&=&\frac{-i}{B_{jk}({\mathbf q}) B_{jl}({\mathbf q})}\nonumber\\
& &\times\{(\omega^2_j({\mathbf q})-\omega^2_k({\mathbf q}))
(\omega^2_j({\mathbf q})-\omega^2_l({\mathbf q}))\nonumber\\
& &-4\gamma^2\omega^2_j({\mathbf q})\}.
\label{I5}
\end{eqnarray}
\begin{eqnarray}
{\cal I}^{(6)}_{jkl}({\mathbf q})
&=&\int^\infty_{-\infty} \frac{d\omega}{2\pi}
\omega^2 D_j({\mathbf q},\omega)D_k(-{\mathbf q},-\omega)
D_l(-{\mathbf q},-\omega) \nonumber\\
&=&\frac{-\gamma}{B_{jk}({\mathbf q})B_{jl}({\mathbf q})}\nonumber\\
& &\times\{(\omega^2_j({\mathbf q})+\omega^2_l({\mathbf q}))
(\omega^2_j({\mathbf q})-\omega^2_k({\mathbf q}))\nonumber\\
& &+2\omega^2_j(\omega^2_j({\mathbf q})-\omega^2_l({\mathbf q}))\}.
\label{I6}
\end{eqnarray}
\begin{equation}
{\cal I}^{(7)}_{jk}({\mathbf q})=\int^\infty_{-\infty} \frac{d\omega}{2\pi}
 \omega D_j({\mathbf q},\omega)D_k({\mathbf q},\omega)= 0. 
\label{I7}
\end{equation}

\section{Wave vector sums in the continuum limit}

Consider the sum 
\begin{equation}
{\cal M}^{(0)}_{jk}=\frac{1}{W_2W_3}\sum_{\mathbf q} 
\frac{1}{B_{jk}(\mathbf q)}
\label{calM0}
\end{equation}
In the continuum limit ,
\begin{equation}
\frac{1}{W_2W_3}\sum_{\mathbf q} \rightarrow \frac{a^2}{(2\pi)^2}
\int^{\frac{\pi}{a}}_{-\frac{\pi}{a}}\int^{\frac{\pi}{a}}_{-\frac{\pi}{a}} 
dq_2dq_3   
\end{equation}
and
\begin{equation}
{\cal M}^{(0)}_{jk} = \frac{16}{\pi^2\omega^4_h}\int^{\frac{\pi}{2}}_0
\int^{\frac{\pi}{2}}_0 \frac{d\theta_2\theta_3}{e_{jk}(\theta_2,\theta_3)},
\end{equation} 
where $\tilde{\gamma}=\frac{2\gamma}{\omega_h}$ and
\begin{equation}
e_{jk}(\theta_2,\theta_3)=E_{jk}+4{\tilde\gamma}^2
(\sin^2\theta_2+\sin^2\theta_3).
\label{ejk}
\end{equation}
$E_{jk}$ is given as
\begin{equation}
E_{jk}=(\cos j\nu-\cos k\nu)^2+
\tilde\gamma^2 (\cos j\nu+\cos k\nu+2).
\label{Ejk}
\end{equation}
We have evaluated the double integrals and expressed the result in terms of 
hypergeometric function as 
\begin{equation}
{\cal M}^{(0)}_{jk}=\frac{4}{\omega^4_h} M^{(0)}_{jk},
\label{calM0M0}
\end{equation}
where 
\begin{eqnarray}
M^{(0)}_{jk} &=&\frac{1}{\Delta_{jk}}F(\frac{1}{2},\frac{1}{2},1;
(\frac{4\tilde\gamma^2}{\Delta_{jk}})^2),
\label{M0}\\
\Delta_{jk} &=& (\cos j\nu -\cos k\nu)^2\nonumber\\
& &+\tilde\gamma^2(\cos j\nu +\cos k\nu +6).
\label{Delta}
\end{eqnarray}
In a similar manner the following sums can also be expressed 
in terms of integrals in the continuum limit:
\begin{eqnarray}
{\cal M}^{(1)}_{jk} &=& \frac{1}{W_2W_3}\sum_{\mathbf q}
\frac{\sin^2{(\frac{q_2a}{2})}}{B_{jk}(\mathbf q)}
= \frac{1}{W_2W_3}\sum_{\mathbf q}
\frac{\sin^2{(\frac{q_3a}{2})}}{B_{jk}(\mathbf q)}
\nonumber\\
&=& \frac{1}{8\gamma^2\omega^2_h}(1-E_{jk}M^{(0)}_{jk}),
\label{calM1}\\
{\cal M}^{(2)}_{jkl} &=& \frac{1}{W_2W_3}\sum_{\mathbf q} 
\frac{1}{B_{jk}(\mathbf q)B_{jl}(\mathbf q)}\nonumber\\
&=& \frac{16}{\omega^8_h}
\frac{M^{(0)}_{jk}-M^{(0)}_{jl}}{E_{jl}-E_{jk}},
\label{calM2}
\end{eqnarray}
The following sums over wave vector are decomposed in terms 
of $M^{(0)}_{jk}$:
\begin{eqnarray}
& &\frac{1}{W_2W_3}\sum_{\mathbf q}\frac{\sin^2{\frac{q_2a}{2}}+
\sin^2{\frac{q_3a}{2}}}{B_{jk}(\mathbf q)B_{jl}(\mathbf q)}\nonumber\\
&=& \frac{1}{\gamma^2\omega^6_h}
\frac{E_{jl}M^{(0)}_{jl}-E_{jk}M^{(0)}_{jk}}{E_{jl}-E_{jk}},
\label{qsum1}\\
& &\frac{1}{W_2W_3}\sum_{\mathbf q}\frac{(\sin^2{\frac{q_2a}{2}}+
\sin^2{\frac{q_3a}{2}})^2}{B_{jk}(\mathbf q)B_{jl}(\mathbf q)}\nonumber\\
&=&\frac{1}{16\gamma^4\omega_h^4}\frac{E_{jl}-E_{jk}+E^2_{jk}M^{(0)}_{jk}
-E^2_{jl}M^{(0)}_{jl}}{E_{jl}-E_{jk}}.
\label{qsum2}
\end{eqnarray}

\section{Definition of useful integrals}

\begin{eqnarray}
M^{(1)}_{jk}&=&\frac{4}{\pi^2}\int_0^{\frac{\pi}{2}}
\int_0^{\frac{\pi}{2}}d\theta_2 d\theta_3
\frac{\sin^2\theta_2+\sin^2\theta_3}{e_{jk}(\theta_2,\theta_3)},
\label{M1}\\
M^{(2)}_{jkl}&=&\frac{4}{\pi^2}\int_0^{\frac{\pi}{2}}
\int_0^{\frac{\pi}{2}}d\theta_2 d\theta_3
\frac{1}{e_{jk}(\theta_2,\theta_3)e_{jl}(\theta_2,\theta_3)},
\label{M2}\\
M^{(3)}_{jkl}&=&\frac{4}{\pi^2}\int_0^{\frac{\pi}{2}}
\int_0^{\frac{\pi}{2}}d\theta_2 d\theta_3
\frac{\sin^2\theta_2+\sin^2\theta_3}
{e_{jk}(\theta_2,\theta_3)e_{jl}(\theta_2,\theta_3)},
\label{M3}\\
M^{(4)}_{jkl}&=&\frac{4}{\pi^2}\int_0^{\frac{\pi}{2}}
\int_0^{\frac{\pi}{2}}d\theta_2 d\theta_3
\frac{(\sin^2\theta_2+\sin^2\theta_3)^2}
{e_{jk}(\theta_2,\theta_3)e_{jl}(\theta_2,\theta_3)}.
\label{M4}
\end{eqnarray}

\section{Discrete sums}

\begin{eqnarray}
& &\sum_{n_1=1}^{N-1} (A_{n_1,j}+A_{n_1+1,j})
(A_{n_1,k}-A_{n_1+1,k}) \nonumber\\
&=&\frac{2}{N+1}(1-(-1)^{j+k})\sin j\nu\sin k\nu\nonumber\\
& &\times\Big(1-\frac{1}{\cos j\nu - \cos k\nu}\Big).
\label{dsum1}
\end{eqnarray}
\begin{eqnarray}
& &{\cal N}(k_1,k_2,k_3,k_4)\nonumber\\
&=& \sum_{n_1=1}^{N-1}(A_{n_1,k_1}+A_{n_1+1,k_1})
(A_{n_1,k_2}-A_{n_1+1,k_2})\nonumber\\
& &\times(A_{n_1,k_3}-A_{n_1+1,k_3})
(A_{n_1,k_4}-A_{n_1+1,k_4})\nonumber\\
&=&\frac{8}{(N+1)^2}\{1-(-1)^{k_1+k_2+k_3+k_4}\}
{\cal N}_1(k_1,k_2,k_3,k_4),\nonumber\\
& &
\label{dsum2}
\end{eqnarray}
where
\begin{eqnarray}
{\cal N}_1(k_1,k_2,k_3,k_4)
&=&\sin k_1\nu\cos\frac{k_2\nu}{2}
\cos\frac{k_3\nu}{2}\cos\frac{k_4\nu}{2}\nonumber\\
& &\times\Big[4\sin\frac{k_2\nu}{2}\sin\frac{k_3\nu}{2}
\sin\frac{k_4\nu}{2}\nonumber\\
& &+\frac{\sin\frac{(k_2-k_3+k_4)\nu}{2}}
{\cos(k_2-k_3+k_4)\nu-\cos k_1\nu}\nonumber\\
& &+\frac{\sin\frac{(k_2+k_3-k_4)\nu}{2}}
{\cos(k_2+k_3-k_4)\nu-\cos k_1\nu}\nonumber\\
& &-\frac{\sin\frac{(k_2-k_3-k_4)\nu}{2}}
{\cos(k_2-k_3-k_4)\nu-\cos k_1\nu}\nonumber\\
& &-\frac{\sin\frac{(k_2+k_3+k_4)\nu}{2}}
{\cos(k_2+k_3+k_4)\nu-\cos k_1\nu}\Big].
\label{calN1}
\end{eqnarray}
${\cal N}(k_1,k_2,k_3,k_4)$ has the following
symmetries:
\begin{eqnarray}
{\cal N}(k_1,k_2,k_3,k_4) &=&
{\cal N}(k_1,k_3,k_2,k_4) 
={\cal N}(k_1,k_4,k_3,k_2)\nonumber\\ 
&=&{\cal N}(k_1,k_2,k_4,k_3).
\label{propcalN}
\end{eqnarray}


\end{document}